# From Interviews to Equations: A Multi-Phase System Dynamics Model of Engineering Student Engagement

Mohammed A. Alrizqi
Sibley School of Mechanical and Aerospace Engineering
Cornell University, Ithaca, NY, USA
Ma867@cornell.edu

**Abstract**

This study presents a method for translating qualitative data into quantitative parameters within a system dynamics (SD) framework to model engineering student engagement. While SD often uses numerical inputs, "soft" factors—motivation, confidence, and sense of belonging—are harder to quantify.

Semi-structured interviews with mechanical engineering students in Learning Studios generated narratives about hands-on coursework, peer support, and growth. Inductive thematic analysis allowed coding of key factors, which were transformed into weighted parameters for a Vensim model. The model comprises interrelated submodels that link community cohesion, motivation, learning outcomes, and aspirations.

Simulations demonstrated exponential growth in motivation, confidence, and belonging, as well as declines in dissatisfaction. A logistic limit on belonging confirmed saturation effects, while a one-week delay across loops slowed but preserved dynamics. These findings are consistent with established educational theory, implying that community-driven interventions can meaningfully enhance student engagement.

This approach underscores the significance of capturing intangible, dynamic student experiences in SD models to design more effective educational interventions. Although further validation with larger samples is warranted, the framework demonstrates how incorporating qualitative insights into quantitative simulations can yield more nuanced and actionable findings for educators and researchers, ultimately informing robust pedagogical and curricular strategies.

# Introduction

**Background**

System Dynamics (SD) has been increasingly applied to educational contexts to analyze complex problems and improve policies. Traditional SD modeling relies on quantitative data, but capturing qualitative factors (e.g., motivation, attitudes) remains challenging. Researchers have long recognized the need to incorporate "soft" variables into SD models, and efforts to do so date back to the 1980s [1], [2], [3], [4], [5]. Over the last decade, several studies have developed methods to bridge qualitative data with formal simulation. For example, Luna-Reyes and Andersen (2003) outlined systematic procedures for collecting and analyzing qualitative data in SD, including interviews and focus groups and analytical techniques like discourse analysis [6]. This laid an important foundation for using rich narrative data to inform system models.

Subsequent research has refined qualitative SD modeling techniques. Halabi et al. (2011) demonstrated one of the first applications of SD to a purely qualitative problem: they conducted semi-structured interviews and coded them to identify key variables and causal links, which were then used to construct causal loop diagrams and stock-and-flow structures [7]. Their study confirmed the feasibility of developing a runnable SD model from interview data, though they noted practical challenges (e.g. some participants declined interviews, and manual coding was time-consuming, prompting use of qualitative analysis software).

Building on such work, Elsawah et al. (2015) introduced a hybrid approach called ICTAM (Interviews, Cognitive Maps, Time-sequence diagrams, All-encompassing conceptual model, then Model) to integrate stakeholder interviews into an agent-based simulation. In this proposed method, the interviews are semi-structured, meaning there is a guiding set of questions with room for interviewee elaboration [8]. Cognitive maps represent the individuals' decision-making processes. The collective map is made by merging individual cognitive maps to model the group's decision-making process. This is followed by a conceptual model meant for transitioning to a numerical model, and the final model, meant for simulation purposes, is the agent-based model of the system. The researchers reported that this can be an applicable method for simulating qualitative systems, although there are potential problems involving researchers getting lost in interviews, gaps between the individual decision-making process and what the individual reveals in interviews, and inaccuracies as a result of personal interpretation and the subjective nature of analyzing interviews. Nevertheless, this paper suggests progress in qualitative system dynamics.

Most recently, Newberry and Carhart (2023) discussed a general method for creating causal loop diagrams based on interviews [9]. Their procedures involve the following steps: coding raw data to identify themes and select relevant interview segments, identifying "microstructures" and causal relationships between variables, converting that into a word and arrow diagram, generalizing into a causal diagram, and generating a "data source reference table" linking back to the interview text. This focus on thorough thematic coding and traceability highlights a wider trend: qualitative modeling in SD is becoming increasingly precise and methodical, allowing for greater confidence in models that depict social or educational phenomena.

In the domain of education, qualitative SD approaches have been used to examine systemic changes and student outcomes. For instance, Tsaple and Tzionas (2019) developed qualitative SD models to assess how Massive Open Online Courses (MOOCs) disrupt traditional higher education [10]. Notably, by creating multiple CLDs from different interviewees, they uncovered diverse perspectives on the MOOC impact. Such work illustrates the value of SD for mapping educational settings qualitatively, especially when quantitative data are scarce. However, many SD models in education still focus on institutional metrics (enrollment, finances, etc.), whereas modeling the soft factors of the student experience, like motivation and engagement, is understudied. This study situates itself at the intersection of these threads: it leverages advances in qualitative SD modeling to capture engineering students' experiential variables within an educational intervention that was implemented by a US University called Learning Studios. These studios are designed to bridge theoretical concepts with practical applications while nurturing a sense of community. The system comprises four essential components: operational engineering systems, advanced analytical tools, streamlined models, and discovery modules aimed at exploring system physics. This setup promotes continuous interaction and experiential learning, allowing students to engage with the same environment from different perspectives and switch between expert and learner roles.

**Objective of this study**

This study aims to construct an empirical system dynamics model that encapsulates the multifaceted effects of learning studios on mechanical engineering students' engagement and experiences. By integrating qualitative data from student interviews within a system dynamics framework, this research seeks to identify and articulate the dynamic interplay between hands-on learning, sense of belonging, personal growth, motivation, and confidence. The objective is to bridge the gap between qualitative perceptions and quantitative analysis, providing a methodological approach for the quantification of qualitative variables in engineering education.

# Methods

A qualitative study grounded in interpretivism was conducted to understand how Learning Studios influence mechanical engineering students' experiences. Interpretivism was chosen to capture the depth and subjective meaning within students' personal narratives. The research design incorporated three key stages: developing an interview protocol guided by the Expectancy-Value-Cost (EVC) theory [11], selecting participants via purposive sampling, and employing semi-structured interviews before engaging in inductive thematic analysis.

The EVC framework—which considers students' success expectations, intrinsic value, attainment value, utility value, and associated costs—shaped the interview protocol. Semi-structured interviews struck a balance between structured prompts and the flexibility to explore emerging topics [12], making them ideal for eliciting rich qualitative data. Eight participants, all mechanical engineering students with Learning Studio experience, were selected to ensure varied perspectives across class standing, gender, and race. Pilot testing refined the interview protocol, and the Institutional Review Board (IRB) approved all procedures.

The interview protocol, data collection, and initial thematic analysis procedures are detailed in our prior work [13], which focused on feedback dynamics within undergraduate engineering education. In the present study, we extend this analysis by applying a multi-phase system dynamic modeling approach to the same qualitative dataset.

Data collection involved approximately 45-minute, audio-recorded interviews conducted in private settings. Analyses followed Braun and Clarke's [14] six-phase thematic process, allowing themes to emerge directly from the data: familiarization, initial coding, theme searching, theme refining, theme defining, and lastly integrating findings into a coherent narrative. This inductive approach ensured that under-researched areas were illuminated by participants' own experiences, thereby enriching our understanding of how Learning Studios influence mechanical engineering students' academic engagement and motivation.

**Causal Loop Diagram Development**

To model the complex dynamic relationships among variables, we constructed a causal loop diagram (CLD) along with a corresponding stock-and-flow model. In System Dynamics, various tools are utilized to analyze the structure of the system, one of which is CLD. CLDs are qualitative models that illustrate the elements and variables within the system while demonstrating the causal relationships among them. Additionally, they offer a straightforward way to observe feedback loops, which can result in complex behaviors [10], [15]

After identifying themes from the interviews, we designed Causal Loop Diagrams (CLDs) to illustrate the feedback dynamics within the students' experiential framework. This method aligns with previous qualitative modeling practices such as Halabi et al. (2011) [7]. They transitioned from coded interview insights to create CLDs, which ultimately led to a simulation model. In our analysis, the interview insights revealed interrelated elements, such as how hands-on projects enhance feelings of competence and enjoyment, which in turn boost motivation and engagement. We documented each link in the CLD with references back to the raw interview statements, akin to the "data source reference table" approach for traceability. This ensures that every relationship in the model (e.g., peer support → sense of belonging, or sense of belonging → reduced need for external validation) is grounded in participant testimony, bolstering the model's credibility.

**Qualitative Variables Quantification**

In our study, we established a detailed framework for quantifying variables within the Vensim model to conduct a robust and meaningful analysis. This framework was designed based on the responses gathered from participants regarding various contributing factors linked to key variables in the study. Meaning if "Factor A" was cited twice as often as "Factor B" as a contributor to some outcome, we assigned a proportionally higher weight to A in the model equations. This approach of using mention frequency as a proxy for influence directly tackles the "uncertainty of values for qualitative variables" that Coyle (1999) highlighted [1]. By transparently converting counts of coded themes into equation parameters, we created a bridge between qualitative insight and quantitative simulation. This methodological choice aligns with broader calls in the literature to find pragmatic ways to quantify qualitative models, and it echoes elements of the ICTAM methodology, where interview-derived cognitive maps were eventually translated into a formal model with quantifiable rules [8].

Below is a summary of the criteria utilized for quantification:

1. Occurrences:
   We logged instances where participants discussed contributing factors to the primary variable. Each mention was recorded separately, measuring the factor's relevance and discussion frequency.

2. Multiple Mentions:

If a participant repeatedly mentioned a contributing factor during their interview, we counted each mention separately. This method captured the emphasis participants placed on specific factors, highlighting their perceived importance and impact.

3. Cumulative Counts Across Participants:
   We aggregated counts from all interviews to determine each factor's overall significance and prevalence within the group.

4. Explicit and Implicit Mentions:
   Our analysis included direct and indirect references, capturing explicit mentions and implied connections for a deeper understanding of how participants link factors to the main variables in their learning experiences.

## Assumption Modelling in System Simulations

In developing the simulation models using Vensim, we established key assumptions to streamline our analysis of the impact of the learning studios on the student's experiences:

### Consistent Hands-On Project Variable

We assume that the influence of hands-on projects on learning remains constant throughout the simulation. This is important as it is the main feature of the learning studios where it was built on real engineering systems applications. This will also simplify the model by stabilizing a core input, allowing me to more clearly observe and analyze the effects of other variables on learning outcomes.

### Impactful Variables on Learning

We have identified the following variables as significant to learning outcomes:

> **Sense of Belonging:** This represents how students perceive their value and inclusion within the learning community. We consider it crucial for fostering engagement and overall academic success.
>
> **Motivation:** We view motivation as both intrinsic and extrinsic drives that influence students' engagement in hands-on projects. It determines their enthusiasm and persistence in overcoming academic challenges.
>
> **Confidence:** Confidence refers to students' self-assurance in their ability to complete tasks and solve problems within the project framework. We expect confidence to grow as students gain positive experiences from project involvement.

These assumptions are vital for model construction, defining the analysis's boundaries. Assuming a constant hands-on project variable simplifies the model but restricts exploration of how project implementation variations impact other variables and learning outcomes.

### Model Formulation

The causal loop diagrams (CLDs) and stocks-flow presented below illustrate the interdependent relationships within learning environments. Those serve as the foundation for our system dynamics modeling. Each submodel focuses on different aspects of the students' experience in the learning studios and their impact on learning. It also highlights the complex interactions that affect the development of the students.

*Submodel 1: Learning and Personal Growth*

This submodel examines the link between personal growth, resilience, and belonging in the learning environment. It suggests that confronting academic challenges fosters resilience and enhances belonging, essential for personal development. As students progress academically, the resilience gained helps their growth, reinforcing a positive feedback loop. The learning process is significantly influenced by practical experiences and a sense of competence, leading to deeper understanding and increased engagement.

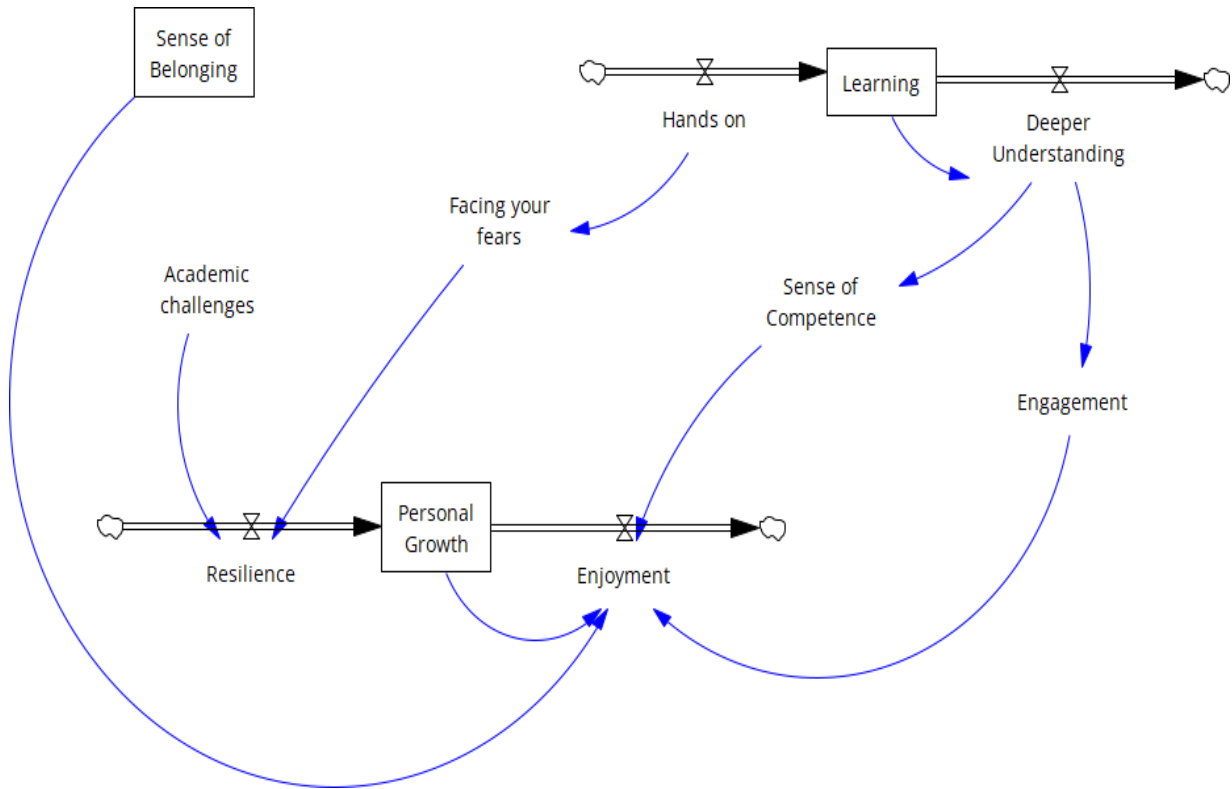

Figure 1. Sub-model 1: Learning and Personal Growth

**Sub model 2: Social Dynamics and Community**

The second sub model emphasizes the social components of the learning environment, such as peer support, a sense of community, and open communication. It demonstrates how a supportive environment can bolster a student's sense of belonging. Additionally, it addresses the impact of competitiveness and impostor syndrome, which can affect students' self-perception and decision-making. Key to this model is the notion that feeling like a 'real engineer' and making the right decisions are influenced by the interconnectedness and support within the student community.

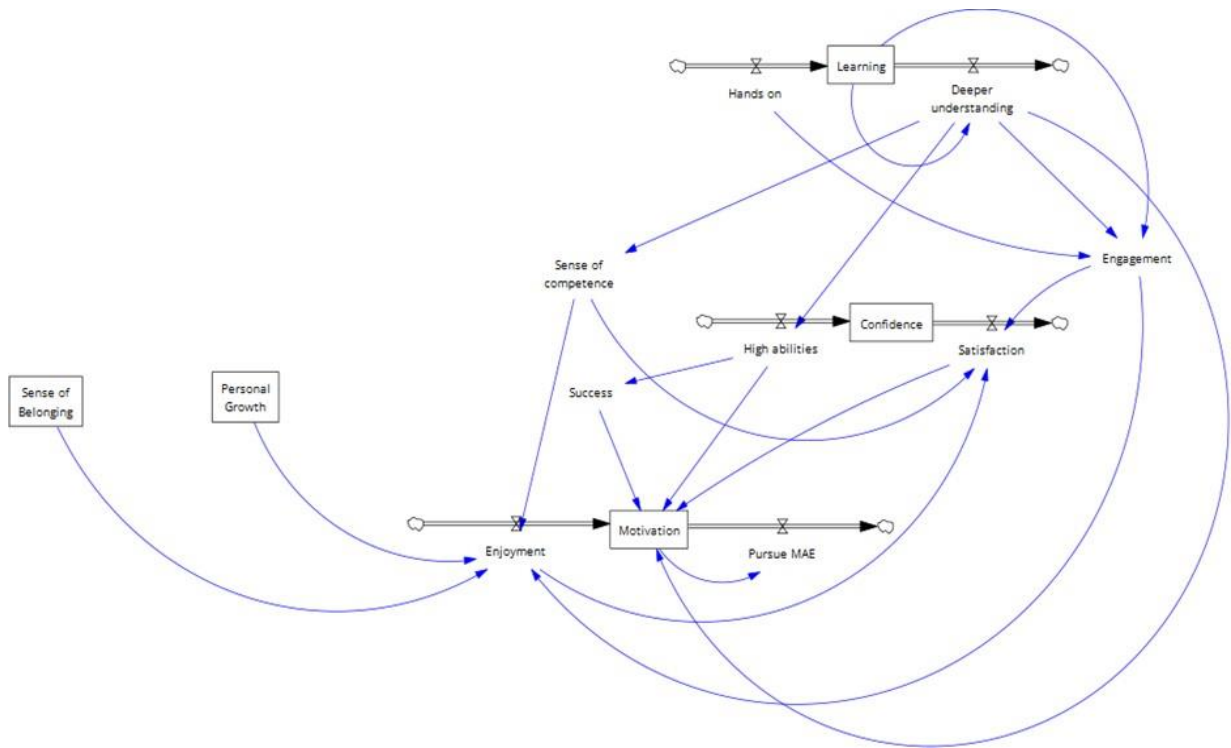

*Figure 2. Sub-model 2: Social Dynamics and Community*

**Submodel 3: Motivation and Career Aspirations**

This submodel emphasizes the motivational components of learning, connecting personal accomplishments and fulfillment to future educational and career aspirations. It illustrates how enjoyment and success in learning activities not only enhance a student's confidence and motivation but may also encourage them to stay in the program, hence increasing the retention rate. This model highlights the critical role of a sense of competence and academic success in fostering overall satisfaction and engagement, which are key drivers for continuing education and professional advancement.

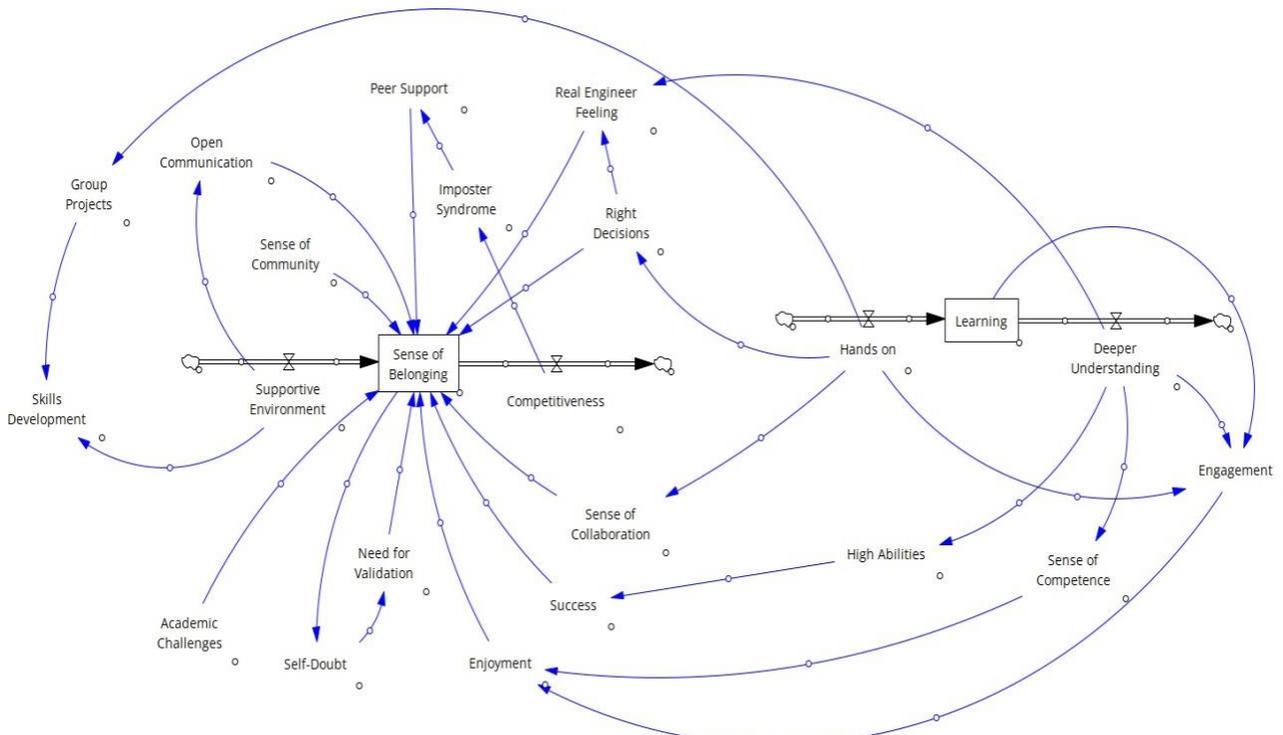

*Figure 3. Sub-model 3: Motivation and Career Aspirations*

# Results and Discussion

Before simulating our model, we needed to assign functions to each variable in the model. These equations were nothing more than a weighted sum of all contributing variables. The weight of each contributing variable was determined based on its frequency relative to the frequency of the other contributing variables. For example, suppose, for variable v, we have contributing variables x, y, and z. X is mentioned 5 times, Y is mentioned 4 times, and Z is mentioned 3 times. That is a total of 12 mentions. Thus, the weight of x is 5/12, the weight of y is 4/12, and the weight of z is 3/12. Altogether, the equation for our variable is $v = (5/12)x + (4/12)y + (3/12)z$. Having assigned equations to all variables, we simulated the model over a period of 16 weeks. We had a total of three simulations: one baseline and two experiments. The baseline simulation was the model we derived it. The results from this simulation are shown below.

## Baseline

Overall, the results show that learning, belonging, confidence, personal growth, and motivation all increase exponentially throughout the 16 weeks. Similarly, negative emotions and detrimental traits, such as dissatisfaction and need for validation, decrease exponentially, indicating that they dwindle away in time. This is a good result as it indicates that learning studios is effective in positively enhancing student experiences.

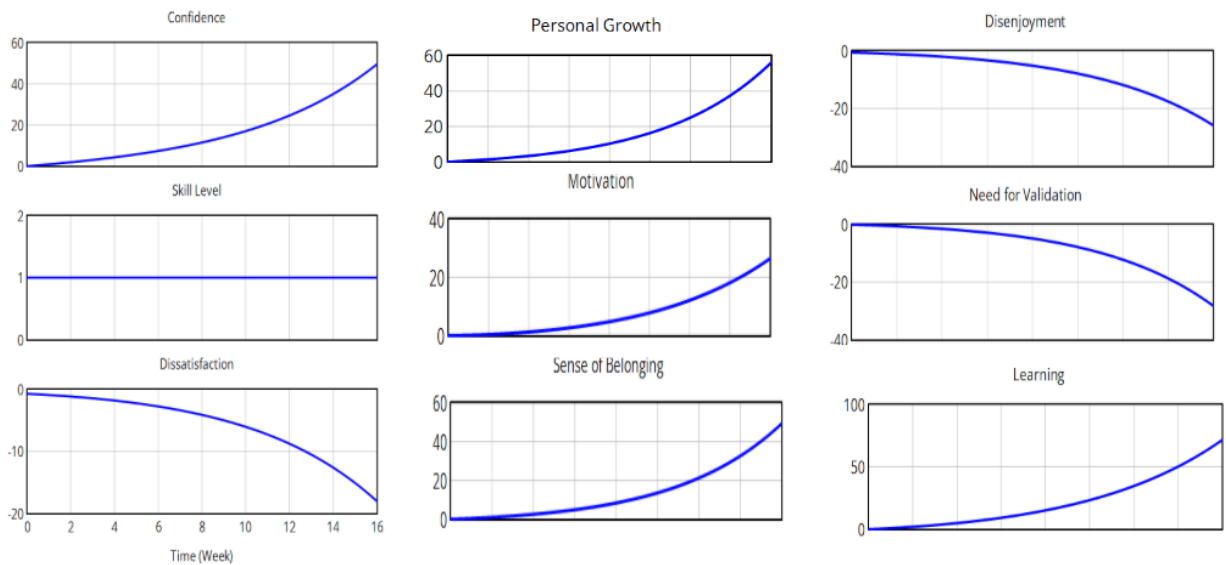

The next scenario is an experiment to see what happens in the model if we place a limit on growth on the sense of belonging. Realistically, one's sense of belonging does not increase forever, so we place a limit on it. This is accomplished in the simulation by implementing a function similar to the logistic equation. Pmax - Pmax / (1 + EXP(-k * (OUR EQUATION- t0)))

## Logistic Limit to Growth on Sense of Belonging

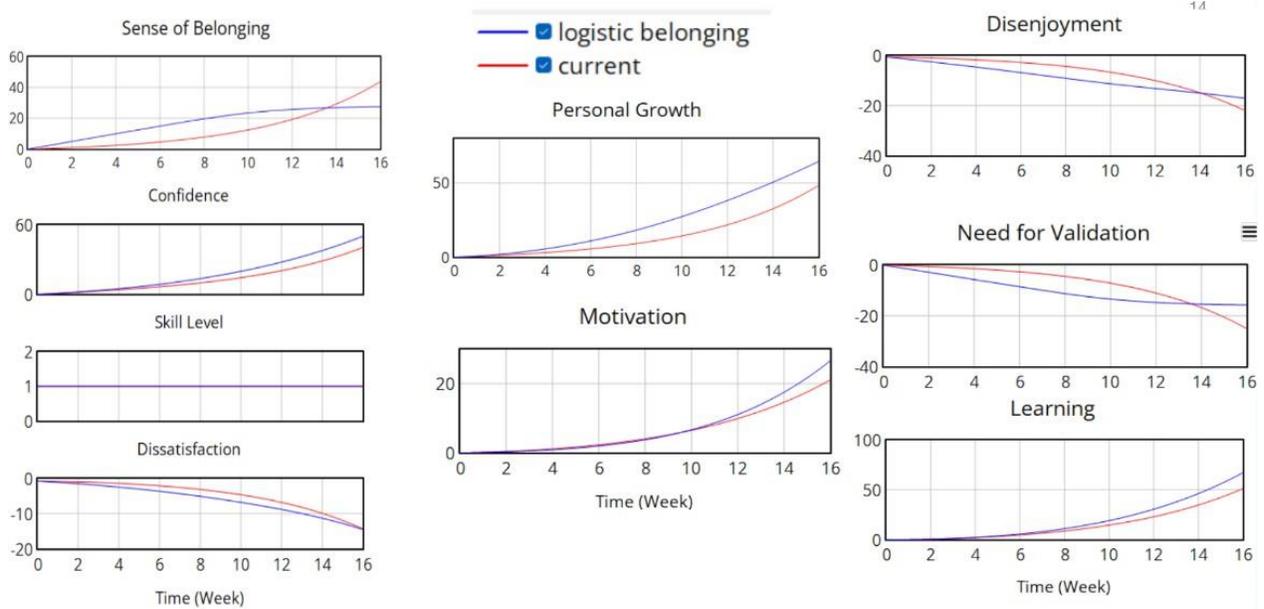

In the graphs pictured here, the red line represents the baseline, and the blue line represents the experimental results. We can see that we successfully created a limit to growth on the sense of belonging. With this change, we also notice that the need for validation decreases, but it reaches a floor, effectively mirroring the behavior in the sense of belonging. Interestingly, we can see that other variable, including confidence, personal growth, motivation, and learning, actually increase faster than in the baseline simulation. It seems that by including this limit to growth (in the interest of incorporating a realistic effect) we actually see even better results than in the baseline simulation.

The next simulation we performed incorporates a delay in the effect of every variable on other variables. Specifically, we altered the model such that the effect of all contributing variables to a given variable is delayed by one week. The results are as follows.

**One-Week Delay**

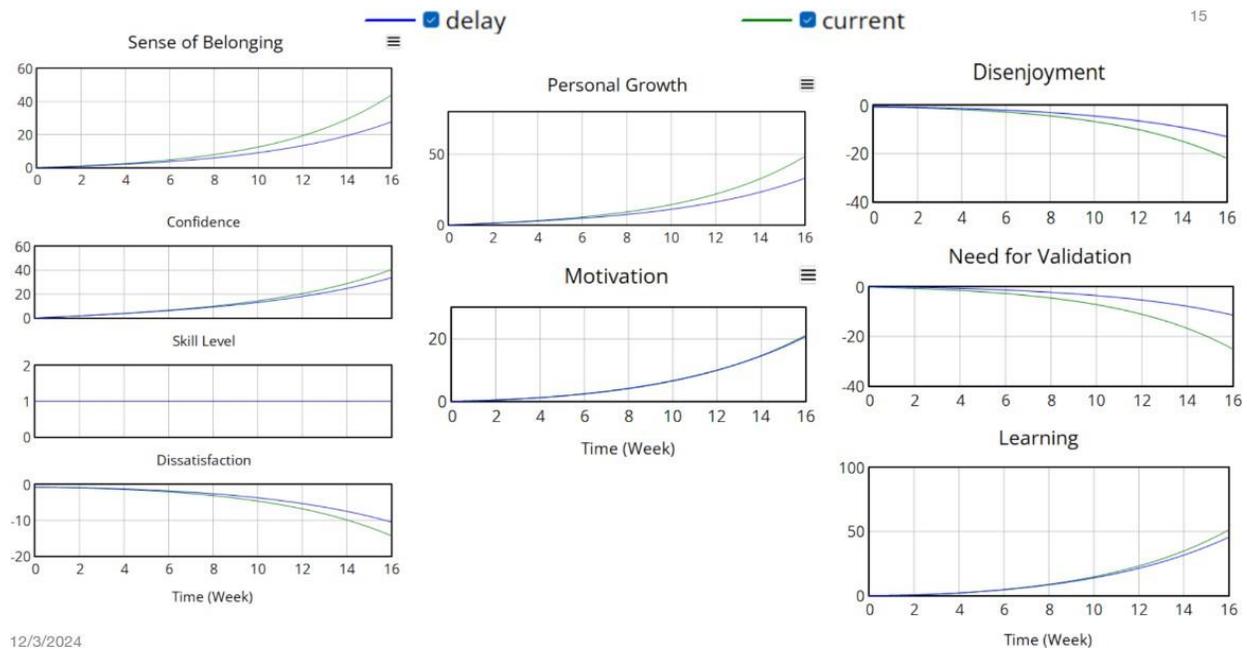

In the results shown, the green line is the baseline model, and the blue line is the delayed model. We see the same patterns of exponential growth and decay that we saw in the baseline model. However, we can see that there is a decreased rate of growth/decay in all variables in comparison to the baseline model. This result makes sense since all effects are delayed, which corresponds to the slower growth rate.

# Discussion

Validating a qualitative simulation is inherently challenging due to the lack of traditional numerical data for comparison. In lieu of direct empirical validation (e.g. time-series data), one important benchmark is consistency with theory and prior findings [16]. Our simulation results showed clear patterns: key positive constructs (learning, motivation, confidence, sense of belonging, etc.) increased over time, while negative feelings (dissatisfaction, need for validation born of insecurity) decreased. We interpret these results by considering established educational theories and research, including the literature on active learning and motivation [17], [18], [19], [20]. Encouragingly, the model's behavior aligns with what one would expect if a learning intervention is truly effective – a notion supported by the literature on active learning and motivation. For instance, it is well-documented that when students feel competent and included, their motivation and engagement rise steadily. Our baseline simulation exhibited exactly this trend, which boosts confidence in the model's validity.

Compared to prior SD models, we see similarities in outcome likelihood. Halabi et al.'s qualitative model, although in a different domain, was deemed useful and feasible by the authors, indicating that a well-crafted qualitative SD model can produce credible dynamics [7]. While they did not have "ground truth" data, the qualitative patterns from their model were considered reasonable representations of the interview insights. Likewise, our model's outputs have face validity when compared with student testimonies and broader educational observations. Additionally, some recent work suggests that if the causal structures are drawn directly from participants' words, the resulting simulations tend to reflect those participants' reality closely [9].

In our study, because each link and variable was grounded in multiple student quotes or repeated mentions, the emergent behavior of the simulation can be viewed as an aggregate narrative of those student experiences [6]. This narrative "rings true" against the backdrop of existing research on learning environments. For example, one of our experimental simulations imposed a realistic limit on the growth of sense of belonging (recognizing that, feelings of belonging would plateau after a certain point). The model responded by showing that beyond that saturation point in belonging, other variables like motivation and learning continued to increase (even accelerating slightly). This outcome is quite intuitive – it suggests that once students reach a healthy level of belonging, it does not hinder further gains in motivation or learning; if anything, it stabilizes social needs and allows focus on growth [18]. Such nuanced behavior adds credibility to the model, as it echoes the psychological understanding that fulfilling belonging can unlock higher achievement. While we must be cautious in interpreting a qualitative simulation, the fact that our results are congruent with both educational theory and empirical studies indicates a form of convergent validity.

In qualitative SD, validation is often about whether the model structure and behavior make sense to experts and stakeholders (in our case, education researchers and the student participants themselves). The alignment we observe – e.g. our model's emphasis on peer support driving belonging is strongly supported by external studies – gives us confidence that the simulation outcomes are not arbitrary. In summary, when positioned against prior qualitative modeling efforts, our study not only showcases a successful application but also implicitly "validates" the approach by producing results that dovetail with real-world expectations and findings.

**Real-World Relevance of Learning Studio Effects**

The model showed the positive impact of learning studios on students' experiences. Studies often show positive student outcomes when projects are included; for example, several studies found that a first-year engineering design studio improved students' community sense and collaboration perception [21], [22]. Students felt more "at home" in the engineering department due to team-based projects [23]. This directly echoes our model, where the sense of community emerged as a crucial factor reinforcing engagement and belonging (with multiple feedback loops tied to it). Moreover, research has highlighted that providing authentic, real-world tasks can sustain student interest even when the novelty wears off. In our simulation experiments, when we introduced a logistic limit on belonging (simulating a realistic leveling-off), we observed that other positive outcomes (motivation, learning gains) continued and even accelerated slightly.

This suggests that once students achieve a comfortable level of belonging and routine in the studio, they might channel more energy into skill development and learning. Real classroom studies similarly note that after an initial adjustment period, students in experiential settings often hit a stride where their focus shifts from "Do I fit in here?" to "What more can I learn and accomplish?". This transition is a desirable educational outcome and may explain why capstone project courses, for example, often see students make significant leaps in ability and confidence toward the end as their comfort in the environment solidifies.

Finally, our findings hold practical implications for broader educational strategies. The fact that our model underscores the importance of peer support, faculty interaction, and hands-on engagement aligns with what higher education institutions are doing to improve student experiences. Many engineering programs now invest in learning communities, mentorship programs, and active learning classrooms to boost belonging and engagement among students. Our study provides a systems perspective on why those investments pay off: it shows how various elements (community, competence, motivation) reinforce each other over time. In a real-world scenario, this means an intervention at one point (e.g. introducing a collaborative project or a faculty mentoring session) can ripple through the system to yield multiple benefits (higher motivation leads to more engagement, which leads to better performance, which further increases motivation, and so on).

This system's view is valuable for educators and administrators designing curricula. It aligns with the notion in research that student experiences are holistic – academic and social factors interconnect – and thus must be addressed together [24]. By situating our results within these established findings, we highlight that our model is not just an abstract simulation, but a reflection of real dynamics observed in educational settings. The learning studio's simulated impact on motivation, engagement, and belonging is strongly backed by empirical studies, reinforcing both the validity of our model and the real-world significance of its insights.

## Limitations

Although this study offers a novel way to convert qualitative data into a system dynamics model, several limitations must be acknowledged. First, the weighting of variables by mention frequency may overemphasize factors reported by particularly vocal participants or underrepresent nuanced but infrequent insights. Second, the assumption of a constant "hands-on project" level simplifies the system but precludes analysis of how varying project intensity might alter students' learning experiences. Third, the lack of direct time-series data means the model relies heavily on theoretical and face-validity checks rather than empirical calibration, leaving space for more robust validation in future research

## Conclusion

This work demonstrates how system dynamics can effectively capture and simulate the complex interaction of motivational and social factors that shape student engagement. By grounding each causal relationship in interview data and systematically converting qualitative themes into weighted parameters, the study offers rich qualitative insights and formal modeling. The simulation results—showing exponential growth in positive constructs like confidence and belonging, alongside a decrease in negative emotions—are consistent with both participant narratives and established educational theory. Furthermore, sensitivity tests with a logistic limit and delayed effects illustrate the utility of SD in exploring realistic constraints and time lags in learning environments. While future inquiries should address the study's limitations and deepen empirical validation, this research underscores the promise of qualitative SD modeling in guiding the design and assessment of impactful educational interventions such as Learning Studios.